\documentclass[twocolumn,aps,floatfix]{revtex4}
\usepackage{graphicx}
\usepackage{amsmath,amsfonts,amssymb}
\usepackage{bm}
\usepackage{natbib}

\begin{document}
\title{Single particle Lagrangian dispersion in Bolgiano turbulence}
\author{A. Bistagnino, G. Boffetta}
\affiliation{Dipartimento di Fisica Generale and INFN,
Universit\`a degli Studi di Torino, Via Pietro Giuria 1,
10125, Torino, \\
and CNR-ISAC, Sezione di Torino, Corso Fiume 4, 10133 Torino, Italy}
\author{A. Mazzino}
\affiliation{CNISM -- Dipartimento di Fisica,
Universit\`a degli Studi di Genova, and Istituto Nazionale di Fisica 
Nucleare, Sezione di Genova, Via Dodecaneso 33, 16146, Genova, Italy}

\date{\today}

\begin{abstract}
Single-particle dispersion in the Bolgiano--Obukhov regime of
two-dimensional turbulent convection is investigated.
Unlike dispersion in a flow displaying the classical K41 phenomenology,
here, the leading contribution to the Lagrangian velocity fluctuations
is given by the largest eddies. This implies a linear behavior
in time for a typical velocity fluctuation in the time interval $t$.
The contribution  to the Lagrangian velocity fluctuations
of local eddies (i.e. with a characteristic time of order $t$),
whose space/time scalings are ruled by the Bolgiano--Obukhov
theory, is thus not detectable
by standard Lagrangian statistical observables. To disentangle 
contributions arising from the large eddies from those of local eddies,
a strategy based on exit-time statistics has successfully been exploited.
Lagrangian velocity increments in Bolgiano convection thus provide 
a physically relevant  example of 
signal with \emph{more than smooth} fluctuations. 

\end{abstract}

\maketitle

%%%%%%%%%%%%%%%%%%%%%%%%%%%%%%%%%%%%%%%%%%%%%%%%%%%%%%%%%%%%%%%%%
Understanding the statistical properties of particle tracers
advected by turbulent flows is a fundamental problem in turbulent
research and a key ingredient for the development of stochastic
models of Lagrangian dispersion \cite{P94,S01,Y02}.
In the recent years there has been great improvements in
theoretical \cite{FGV01}, experimental \cite{LVCAB01,MMMP01,OM00} 
and numerical \cite{Y01,BBCLT05} understanding of Lagrangian turbulence.
Most of the studies have been concentrated on the so-called
single particle dispersion in which the statistical objects
are the Lagrangian velocity differences following a single
trajectory $\delta {\bm v}(t)={\bm v}(t)-{\bm v}(0)$.
For this quantity, dimensional analysis in statistical stationary,
homogeneous and isotropic turbulence, predicts
$\langle \delta v^2 \rangle \simeq \varepsilon t$
(where $\varepsilon$ is the mean energy dissipation).
The remarkable coincidence with diffusive behavior is at the
basis of stochastic models of turbulent dispersion.

In this Letter we investigate on the basis of Direct Numerical Simulations
(DNS) the statistics of single particle dispersion in two-dimensional
Boussinesq convective
turbulence. This system is characterized by an
inverse cascade with scaling exponent in agreement with the 
Bolgiano-Obukhov theory of turbulent convection \cite{S94}. As a consequence,
we expect a non-diffusive behavior for Lagrangian velocity variance.
We show that a careful statistical analysis, based on exit time 
statistics, is necessary in order to observe the expected scaling.

The two-dimensional Boussinesq turbulent convection is described by the 
following set of partial differential equations \cite{CMV01}:

\begin{eqnarray}
\partial_t\omega+{\bm v}\cdot{\bm \nabla}\omega &=&
+\nu\Delta\omega-\beta{\bm \nabla}T\times{\bm g} \nonumber \\
\partial_t T+{\bm v}\cdot{\bm \nabla}T &=&
\kappa\Delta T
\label{eq:1}
\end{eqnarray}
where $T$ is the temperature field and $\omega={\bm \nabla}\times{\bm v}$
is the scalar vorticity, ${\bm g}$ is the gravitational acceleration,
$\beta$ is the thermal expansion coefficient and $\kappa$ and $\nu$ are
molecular diffusivity and viscosity.
Energy in (\ref{eq:1}) is injected by maintaining a mean temperature
profile $\langle T({\bm r},t)\rangle={\bm G}\cdot{\bm r}$,
with a constant gradient ${\bm G}$ pointing in the direction of gravity. 

%%%%%%%%%%%%%%%%%%%%%%%%%%%%%%%%%%%%%%%%%%%%%%%%%%%%%%%%%%%%%%%%%
\begin{figure}[htb!]
\includegraphics[width=0.49\textwidth]{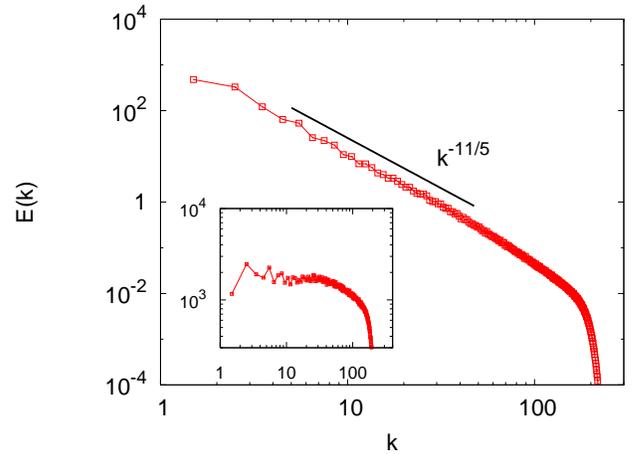}
\caption{Energy spectrum from a Direct Numerical Simulations
of Boussinesq equations at resolution $N=1024^2$ in stationary 
conditions. The line represents the dimensional prediction 
$E(k) \simeq k^{-11/5}$.
The inset shows the spectrum compensated with the prediction.}
\label{fig:spv} 
\end{figure}
%%%%%%%%%%%%%%%%%%%%%%%%%%%%%%%%%%%%%%%%%%%%%%%%%%%%%%%%%%%%%%%%%

In equation~(\ref{eq:1}) the temperature field affects
the vorticity through the buoyancy force, thus providing an
example of {\em active} tracers. At large enough
values of $\beta$, the buoyancy force can equilibrate the inertial
terms in the velocity dynamics, thus providing the mechanism 
at the basis of the Bolgiano-Obukhov theory of turbulent convection.

Let us briefly recall the phenomenology of two-dimensional 
turbulent convection.
The balance of buoyancy and inertial terms in equation~(\ref{eq:1})
introduces the Bolgiano length scale $l_{\rm B}$ \cite{MY}.
At small scales, $r\ll l_{\rm B}$, the inertial term is larger than
buoyancy force and the temperature is basically a passive scalar. At
scales $r\gg l_{\rm B}$, buoyancy dominates and thus affects the velocity
field in the inverse energy cascade. At variance with Navier-Stokes 
turbulence, kinetic energy is here injected at all 
scales in the inverse cascade. The energy flux at scale $r$ is obtained
by a dimensional argument as:
\begin{equation}
\varepsilon_r = \beta{\bm g}\cdot\langle{\bm v}
({\bm x}+{\bm r},t)T({\bm x},t)\rangle
\sim r^{4/5}
\label{eq:2}
\end{equation}
which gives the dimensional prediction for the velocity structure functions:
\begin{equation}
S_p(r)=\langle\left(\delta_r v\right)^p\rangle
\sim \left( \epsilon_r r \right)^{p/3} \sim r^{\zeta_p}
\label{eq:3}
\end{equation}
with $\zeta_p=3p/5$.
The exponent of the second-order structure function ($\zeta_2=6/5$)
gives the scaling exponent for the energy spectrum 
$E(k) \simeq k^{-11/5}$ which is shown in Fig.~\ref{fig:spv}.
It is worth remarking that detailed numerical simulations have
shown that velocity fluctuations display self-similar statistics 
without intermittency corrections \cite{CMV01}.

The dimensional prediction for the statistics of Lagrangian velocity
increments $\delta_t v=v(t)-v(0)$ is the following.
Considering the velocity $v$ as the superposition of the contributions
from eddies of different sizes, the variation $\delta_t v$ over a time
$t$ will be given by the superposition of the variations associated 
to the eddies.
The eddies at scale $r$, with a characteristic time $\tau_r \ll t$
will be decorrelated and thus give no contribution.
The eddies for which $\tau_r \simeq t$ are the first ones to be considered. 
Assuming a scaling exponent $h$ for the velocity the characteristic 
turnover time time is $\tau_r \simeq \tau_L (r/L)^{1-h}$ ($\tau_L$ being
characteristic time of eddies at the large scale $L$)
and thus the scale of the eddies which decorrelates in a time $t$ is given by
$r \simeq L (t/\tau_L)^{1/(1-h)}$. 
Their contribution to the Lagrangian velocity 
fluctuation is thus estimated as
\begin{equation}
\delta_t v \simeq  v_L (r/L)^h \simeq v_L (t/\tau_L)^{q}
\label{eq:4}
\end{equation}
where $v_L$ represents velocity fluctuations at large scales
and $q \equiv h/(1-h)$.

On top of this, the eddies at the large scales of the order of  $L$ 
have to be considered. 
At these scales $t \ll \tau_L$ and the contribution to velocity fluctuations
is differentiable, i.e. $\delta_t v \simeq (\partial_t v_L) t$.
The typical velocity fluctuation on the time interval $t$ is then given
by the superposition of two contributions:
\begin{equation} 
\delta_t v  \simeq  \tau_L (\partial_t v_L) (t/\tau_L) + v_L (t/\tau_L)^q  .
\label{eq:5}
\end{equation}

\begin{figure}[htb!]
\includegraphics[angle=0, width=0.49\textwidth]{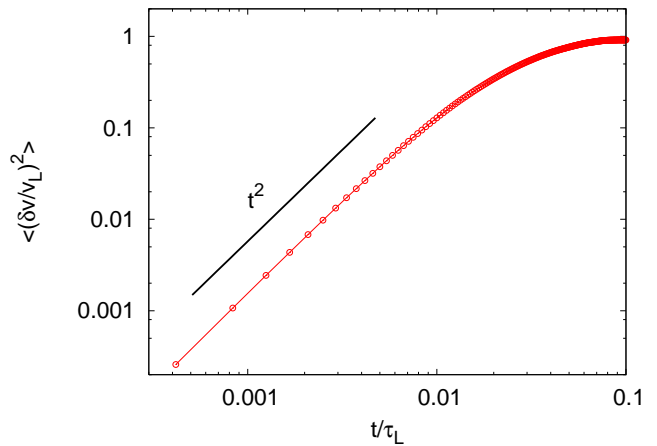}
\caption{Second order Lagrangian structure function. The dominant
ballistic regime ($\delta_t v \simeq t$) is evident.}
\label{fig2}
\end{figure}

At small $t/\tau_L$ the dominant term will be the one with minimum exponent:
$\delta v \sim \min(1,q)$.
In the framework of the classical K41 theory ($h=1/3$, $q=1/2$) the dominant 
contribution in (\ref{eq:5}) is the local one which leads to the 
diffusive behavior $\delta_t v \sim t^{1/2}$.
In the present case of Bolgiano convection we have $q=3/2$ and thus
velocity increments in the inertial range are dominated by the infrared 
term $\propto t$.
Therefore, a standard analysis of velocity fluctuations, i.e. Lagrangian
structure functions $S^{L}_p(t)=\langle \left( \delta_t v \right)^p \rangle$,
is unable to disentangle the Bolgiano-Obukhov scaling in the 
Lagrangian statistics (see Fig.~\ref{fig2}).

Lagrangian velocity increments in Bolgiano convection provide 
a physically relevant  example of 
signal with \emph{more than smooth} fluctuations. 
The statistical analysis of this kind of signal has been recently addressed
on the basis of an {\em exit-time statistics} \cite{BCLVV01}.
In the following we will discuss the application of this approach to
the present case and the resulting bifractal distribution of scaling
exponents.

%%%%%%%%%%%%%%%%%%%%%%%%%%%%%%%%%%%%%%%%%%%%%%%%%%%%%%%%%%%%%%%%%%
%%%%%%%%%%%%%%%%%%%%%%%%%%%%%%%%%%%%%%%%%%%%%%%%%%%%%%%%%%%%%%%%%%

The exit-time statistics is based on the time increments $T(\delta v)$
needed for a tracer to observe a change of $\delta v$ is its
velocity \cite{abccv97}. Given the set of thresholds
$\delta v_n = \rho^n \delta v(0)$ (with $\rho>1$), one computes the
exit-times $T_n$ corresponding to each threshold following the tracer
trajectories. Now, let us assume to have a signal $\delta v$ composed
by two contributions as in (\ref{eq:5}). In the limit of small 
$\delta v$, the differentiable part $\propto t$ will dominate
except when the derivative $\partial_t v_L$ vanishes and the 
local part thus becomes the leading one. For a signal with $1 \leq q \leq 2$,
its first derivative is a one-dimensional self-affine signal with 
H\"older exponent $\xi = q-1$, which thus vanishes on a fractal set 
of dimension $D=1-\xi=2-q$.
Therefore, the probability to observe the component $O(t^q)$ is
equal to the probability to pick a point on the fractal set of dimension $D$,
i.e.:
\begin{equation}
P(T \sim \delta v^{1/q}) \sim T^{1-D} \sim (\delta v)^{1-1/q} \; .
\label{eq:6}
\end{equation}
By using this probability for computing the average $p$-order moments 
of exit-time statistics one obtains the following bi-fractal prediction
\cite{BCLVV01}
\begin{equation}
\langle T^{p}(\delta v) \rangle \sim \delta v^{\chi(p)} \;, \;\; 
\mbox{with} \;\; \chi(p)=\min(p,\frac{p}{q}+1-\frac{1}{q}) \; .
\label{eq:7}
\end{equation}
According to prediction \eqref{eq:7}, low-order moments ($p \leq 1$) of the
inverse statistics only see the differentiable part of the signal,
while high-order moments ($p \geq 1$) are dominated by the local
fluctuations $O(t^q)$.

%%%%%%%%%%%%%%%%%%%%%%%%%%%%%%%%%%%%%%%%%%%%%%%%%%%%%%%%%%%%%%%%%%
\begin{figure}[htb!]
\includegraphics[angle=0, width=0.49\textwidth]{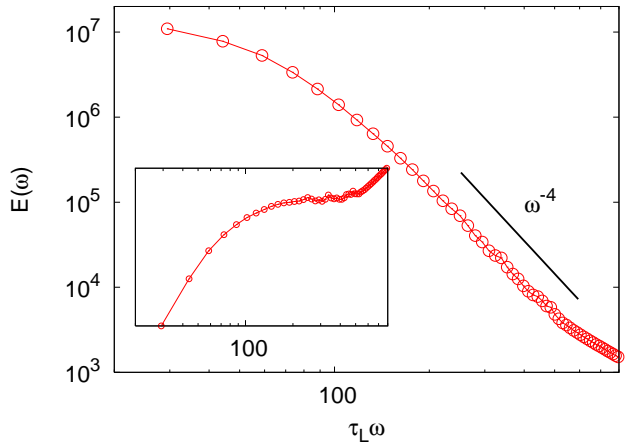}
\caption{The spectrum $E(\omega)$ of the Lagrangian velocity increments 
$\delta v$. The line represent the prediction $\omega^{-4}$.
The inset shows the compensated plot $E(\omega)/\omega^{-4}$. The
$\omega^{-4}$ contribution is due to the superimposed \emph{more than smooth}
fluctuations, while the tail of the spectrum are dominated by the
contributions due to the non-periodicity of the signal.}
\label{fig3}
\end{figure}
%%%%%%%%%%%%%%%%%%%%%%%%%%%%%%%%%%%%%%%%%%%%%%%%%%%%%%%%%%%%%%%%%%

We now turn to the numerical procedure. The turbulent velocity field is
obtained by direct integration of convective Navier--Stokes equation
\eqref{eq:1} with a standard, fully-dealiased pseudospectral method in 
a doubly periodic square domain of resolution $1024^2$. 
For numerical convenience, the viscous term in \eqref{eq:1} has been
replaced in our simulations by a hyperviscous term of order eight.
Time evolution is implemented by a standard second-order
Runge--Kutta scheme. 
Equations are integrated for about one hundred of large-scale
eddy turnover times to reach a stationary state.
Lagrangian trajectories are then obtained by integrating 
${\bf \dot{x}}(t)={\bm v}( {\bm x}(t),t)$ with the
velocity at particle position obtained by linear interpolation from the
nearest grid points. A single run follows 64000 particles, 
homogeneously distributed on the integration domain.
Velocity variations $\delta_t v$ and exit times $T(\delta v)$ of single 
particles are recorded for times comparable to the integral scale.
Average is performed over about 150 runs.

In Fig.~\ref{fig2} we show the variance of Lagrangian velocity fluctuations
as a function of time. 
The smooth behavior 
$\langle (\delta_t v)^2 \rangle \sim t^2$ is clearly observable
making impossible the observation of the Bolgiano-Obukhov scaling.
We remark that the same result is expected for any system with 
large scale domination over the local contribution, i.e. for any
$q \le 1$. Therefore, for these systems (with $h \le 1/2$) the 
direct computation of Lagrangian structure functions (\ref{eq:3}) 
is unable to disentangle the turbulent components in the velocity field.

%%%%%%%%%%%%%%%%%%%%%%%%%%%%%%%%%%%%%%%%%%%%%%%%%%%%%%%%%%%%%%%%%%%
\begin{figure}[htb!]
\includegraphics[width=0.49\textwidth]{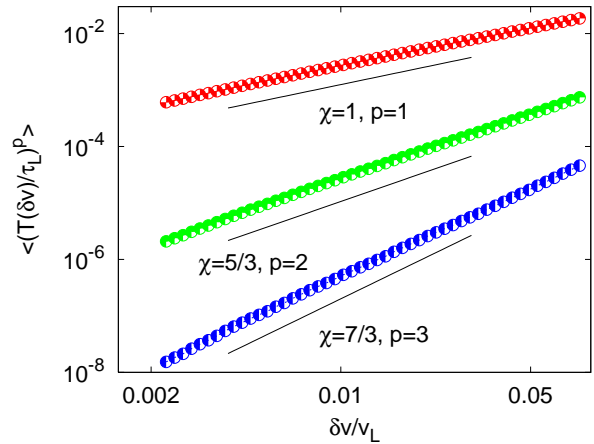}
\caption{Exit times $\langle T^{p}(\delta v) \rangle$ of
of order $p=1,2,3$ vs $\delta v$. Lines represent the predicted 
power-law behavior \eqref{eq:7} with exponents $\chi(p)$.}
\label{fig4} 
\end{figure}
%%%%%%%%%%%%%%%%%%%%%%%%%%%%%%%%%%%%%%%%%%%%%%%%%%%%%%%%%%%%%%%%%%%

A first check of Bolgiano scaling in Lagrangian statistics is
obtained by computing the temporal energy spectrum $E(\omega)$.
We remind that for an infinitely differentiable signal, the smooth
contribution gives an exponential spectrum, while the 
$\delta_t v \sim t^{3/2}$ contribution gives a spectrum proportional to 
$\omega^{-4}$. Figure~\ref{fig3} clearly shows the expected 
$\omega^{-4}$ behavior for an intermediate range of frequencies. 
The high frequencies are dominated by the effects of non-periodicity
of the signal.

\begin{figure}[htb!]
\includegraphics[width=0.49\textwidth]{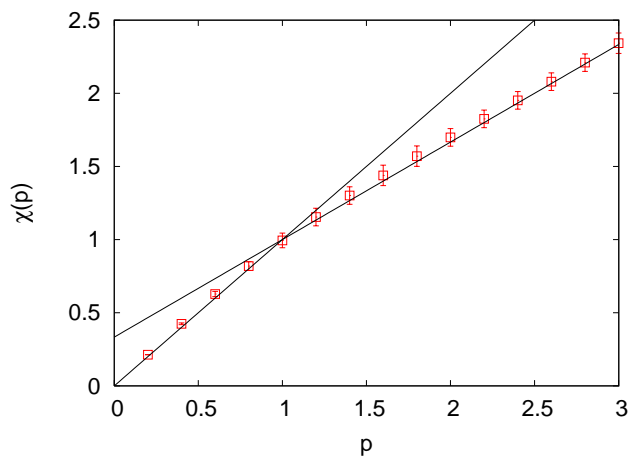}
\caption{Scaling exponents $\chi(p)$ for the moments of
the exit times $\langle T^{p}(\delta v) \rangle$.
The lines represent the bifractal prediction \eqref{eq:7}.
The error bars on the exponents have been estimated by evaluating
differences in $\chi(p)$ changing the fitting interval.}
\label{fig5}
\end{figure}

We now turn to the exit-time analysis. 
Figure~\ref{fig4} shows the first moments of exit times 
$\langle T^p(\delta v) \rangle$ which display a clear power-law
scaling in the range 
$2 \times 10^{-3} v_L \le \delta v \le 5 \times 10^{-2} v_L$.
We were able to compute the moments up to $p=3$ with statistical
significance. The set of scaling exponents obtained from a best fit 
in this range is shown in Fig.~\ref{fig5}. The bifractal spectrum predicted
by (\ref{eq:7}) is clearly reproduced. We remark that the 
fact that for $p>1$ exponents follows the linear behavior 
$\chi(p)=(2 p +1)/3$ indicates the absence of intermittency
in Lagrangian statistics. This feature is a consequence of the 
self-similarity of the inverse cascade in two-dimensional Bolgiano
convection.

%%%%%%%%%%%%%%%%%%%%%%%%%%%%%%%%%%%%%%%%%%%%%%%%%%%%%%%%%%%%%%%%%%%
%%%%%%%%%%%%%%%%%%%%%%%%%%%%%%%%%%%%%%%%%%%%%%%%%%%%%%%%%%%%%%%%%%%

%%%%%%%%%%%%%%%%%%%%%%%%%%%%%%%%%%%%%%%%%%%%%%%%%%%%%%%%%%%%%%%%%
In conclusions, we have analyzed single-particle dispersion in the 
two-dimensional Bolgiano turbulent convection. A remarkable property of this
turbulent regime is that the single-particle dispersion turns out to be
dominated by the large-scale eddies whose contribution to velocity fluctuations
behaves linearly with time. A completely different scenario thus emerge
with respect to the classical single-particle dispersion in a flow
which displays   the standard K41 phenomenology. 
The Bolgiano--Obukhov contribution to the scaling of Lagrangian velocity 
increments thus results undetectable by  standard Lagrangian structure
functions. To disentangle its effect from the (trivial) 
one played by large scales, 
we have  exploited  a method of analysis based on exit-time
statistics for signals having more than smooth fluctuations.
The obtained results clearly permit to identify  
the contribution of ''local'' eddies
to the Lagrangian velocity fluctuations.

%%%%%%%%%%%%%%%%%%%%%%%%%%%%%%%%%%%%%%%%%%%%%%%%%%%%%%%%%%%%%%%%%%%%%%%%%%%%
\vspace{5mm}
This work has been supported by COFIN 2005 project
n.~2005027808 and by CINFAI consortium (A.M.).

%%%%%%%%%%%%%%%%%%%%%%%%%%%%%%%%%%%%%%%%%%%%%%%%%%%%%%%%%%%%%%%%%%%%%%%%%%%%

%%%%%%%%%%%%%%%%%%%%%%%%%%%%%%%%%%%%%%%%%%%%%%%%%%%%%%%%%%%%%%%%%

\end{document}